\documentclass[12pt]{article}

\usepackage{amssymb}
\usepackage{amsmath}
\usepackage{amscd}
\usepackage{latexsym}
\usepackage{graphicx}

\usepackage{cite}

\newcommand{\be}{\begin{equation}}
\newcommand{\ee}{\end{equation}}

\newcommand{\dlt}{\delta}
\newcommand{\prt}{\partial}
\newcommand{\br}{{\bf r}}
\newcommand{\bk}{{\bf k}}

\newcommand{\ep}{\varepsilon}

\newcommand{\ra}{\rightarrow}
\newcommand{\sgm}{\sigma}

\newcommand{\gm}{\gamma}

\newcommand{\Om}{\Omega}

\newcommand{\dgr}{\dagger}
\newcommand{\lbd}{\lambda}
\newcommand{\Lbd}{\Lambda}
\newcommand{\rgl}{\rangle}
\newcommand{\lgl}{\langle}

\begin{document}

\begin{center}

{\Large{\bf Hartree-Fock-Bogolubov method in the theory of 
Bose-condensed systems} \\ [5mm]

V.I. Yukalov$^{1,2}$ and E.P. Yukalova$^{3}$ }  \\ [3mm]

{\it
$^1$Bogolubov Laboratory of Theoretical Physics, \\
Joint Institute for Nuclear Research, Dubna 141980, Russia \\ [2mm]

$^2$Instituto de Fisica de S\~ao Carlos, Universidade de S\~ao Paulo, \\
CP 369,  S\~ao Carlos 13560-970, S\~ao Paulo, Brazil \\ [2mm]

$^3$Laboratory of Information Technologies, \\
Joint Institute for Nuclear Research, Dubna 141980, Russia } \\ [3mm]

{\bf E-mails}: {\it yukalov@theor.jinr.ru}, ~~ {\it yukalova@theor.jinr.ru}

\end{center}

\vskip 1cm

\begin{abstract}
The Hohenberg-Martin dilemma of conserving versus gapless theories for systems with
Bose-Einstein condensate is considered. This dilemma states that, generally, a theory
characterizing a system with broken global gauge symmetry, which is necessary for 
Bose-Einstein condensation, is either conserving, but has a gap in its spectrum, or 
is gapless, but does not obey conservation laws. In other words, such a system either 
displays a gapless spectrum, which is necessary for condensate existence, but is not 
conserving, which implies that it corresponds to an unstable system, or it respects 
conservation laws, describing a stable system, but the spectrum acquires a gap, which 
means that the condensate cannot appear. An approach is described, resolving this 
dilemma, and it is shown to give good quantitative agreement with experimental data. 
Calculations are accomplished in the Hartree-Fock-Bogolubov approximation.   
\end{abstract}

\vskip 1.5cm

\section{Condensate existence and stability}

The necessary and sufficient condition for Bose-Einstein condensation is the spontaneous
breaking of global gauge symmetry $U(1)$ (see \cite{Lieb_1,Yukalov_2,Yukalov_3}). This 
implies that below the Bose condensation temperature $T_c$ the statistical average of a 
field operator $\langle \hat{\psi} \rangle$ becomes nonzero, since it is proportional
to $\sqrt{\rho_0}$, where $\rho_0 \equiv N_0/V$ is the average condensate density. According
to the Bogolubov-Ginibre theorem \cite{Bogolubov_4,Bogolubov_5,Ginibre_6} the condensate 
density $\rho_0$ is a minimizer of thermodynamic potential. The same concerns the density
of uncondensed particles $\rho_1$ that can be expressed through $\rho_0$ as 
$\rho_1 = \rho - \rho_0$, with $\rho$ being a fixed average particle density. Keeping this 
in mind, it is possible to formulate the conditions of condensate existence and stability.
Below, we use the system of units, where the Planck and Boltzmann constants are set to one. 

The general criterion of Bose-Einstein condensation is the occurrence of a nonzero 
thermodynamic limit
\be
\label{1}
\lim_{V\ra\infty} \; \frac{N_0}{V} = \rho_0 > 0 \;   .
\ee
The number of condensed particles $N_0$ can be defined \cite{Penrose_7} as the largest
eigenvalue of the single-particle density matrix. Denoting the set of these eigenvalues
through $n_k$, with $k$ labelling the single-particle states, we have $N_0 = \sup_k n_k$.  
Then limit (\ref{1}) implies that 
\be
\label{2}
 \sup_k n_k \; \propto \; N \ra \infty \qquad ( N \ra \infty) \; .
\ee
The distribution $n_k = n_k(\varepsilon_k)$ is a function of the spectrum of collective 
excitations $\varepsilon_k$. To have the above singularity in the thermodynamic limit, 
there should exist an index $k_0$, such that $n_k$ would diverge, when $k \ra k_0$. 
Respectively \cite{Yukalov_3}, for this index, the spectrum has to tend to zero,
\be
\label{3}    
\ep_k \ra 0 \qquad ( k \ra k_0 ) \;  ,
\ee
which means that the spectrum is gapless. For a uniform system, $k$ represents momentum, 
while $k_0$ becomes zero.  

In the case of a uniform system, the gapless spectrum results \cite{Bogolubov_5,Hugenholtz_8}
in the first expression for the chemical potential    
\be
\label{4}
\mu_1 = \Sigma_{11}(0,0) - \Sigma_{12}(0,0) \;  ,
\ee
called the Hugenholtz-Pines relation, where $\Sigma_{11}({\bf k},\omega)$ is the normal 
self-energy and $\Sigma_{12}({\bf k},\omega)$ is the anomalous self-energy.  
 
The system stability requires that the condensate density be the minimizer of thermodynamic 
potential. This implies for the grand potential the condition
\be
\label{5}
 \frac{\prt\Om}{\prt N_0} = 0 \; ,
\ee
which gives \cite{Yukalov_3} the second expression for the chemical potential
\be
\label{6}
\mu_0 = \Sigma_{11}(0,0) + \Sigma_{12}(0,0) - 2\rho_0 \Phi_0 \;   ,
\ee
where $\Phi_0$ is the Fourier transform of the interaction potential at zero momentum.
As is evident, these two expressions $\mu_0$ and $\mu_1$, generally, do not coincide with
each other. They may coincide in some approximations (to be considered below) or when 
$\Sigma_{12}(0,0) = 0$ and $\rho_0 = 0$, if particle interactions are not neglected. But
the latter simply means that there is no Bose condensate at al. It is possible to show 
that equating $\mu_0$ and $\mu_1$, one comes to the conclusion \cite{Nepomnyashchii_9}
that $\Sigma_{12}(0,0) = 0$.  

Thus we confront the problem of accepting for the chemical potential either $\mu_0$ or 
$\mu_1$. However, if we accept $\mu_1$, then the condensate could exist, but it can be 
not stable. While, if we accept $\mu_0$, the condensate could be stable, but it may 
not exist. This is, actually, the meaning of the Hohenberg-Martin dilemma 
\cite{Hohenberg_9}.

It is important to stress that there is no any thermodynamic requirement that the two 
expressions for the chemical potential be equal \cite{Yukalov_10,Yukalov_11}. Thus, from 
the variation of free energy the equality follows $(\mu_0-\mu_1)\dlt N_0=0$. It is only 
if $N_0$ would be undefined and could be freely varied, the chemical potentials should 
be equal. However, the number of condensed particles is prescribed by the conditions 
of the condensate existence and stability, being uniquely defined for fixed temperature 
and density. Hence $\dlt N_0 = 0$, because of which there is no any constraint on the 
chemical potentials. 
 
As an illustration of the possibility that the condensate could formally exist, 
being in reality unstable, let us consider a uniform ideal Bose gas in the space of 
dimensionality $d$. Then, below the condensation temperature
\be
\label{7}
 T_c = \frac{2\pi}{m} \left[ \frac{\rho}{g_{1/2}(1)} \right]^{1/d} \; ,
\ee  
where $g_n$ is the modified Bose function \cite{Yukalov_11,Yukalov_12}, formally there occurs 
Bose-Einstein condensation, when the condensate density becomes nonzero for any dimensionality 
$d$. However, the stability condition for particle fluctuations
\be
\label{8a}
 0 \leq \frac{{\rm var}(\hat N)}{N} < \infty \qquad 
\left( \; {\rm var}(\hat N) \equiv \lgl \hat N^2\rgl - \lgl \hat N \rgl^2 \; \right)  
\ee
is valid not for all dimensionalities. It is possible to show \cite{Yukalov_13} that, 
depending on spatial dimensionality, particle fluctuations scale as
$$
\frac{{\rm var}(\hat N)}{N} \; \propto \; N^3 \quad ( d = 1 ) \; ;
\qquad
\frac{{\rm var}(\hat N)}{N} \; \propto \; N \quad ( d = 2 ) \; ;
\qquad
\frac{{\rm var}(\hat N)}{N} \; \propto \; N^{1/3} \quad ( d = 3 ) \; ;
$$
$$
\frac{{\rm var}(\hat N)}{N} \; \propto \; \ln N \quad ( d = 4 ) \; ;
\qquad
\frac{{\rm var}(\hat N)}{N} \; \propto \; const \quad ( d > 4 ) \; .
$$
This tells us that the ideal system with a Bose-Einstein condensate can be stable only 
for $d > 4$.

\section{General self-consistent approach}

Here, we describe a general self-consistent approach resolving the Hohenberg-Martin 
dilemma. The Hamiltonian, in the second-quantized picture, is
$$
\hat H = \int \hat\psi^\dgr(\br) 
\left( -\; \frac{\nabla^2}{2m} + U \right) \hat\psi(\br)\; d\br \; +
$$
\be
\label{8b}
+ \; \frac{1}{2} \int \hat\psi^\dgr(\br) \hat\psi^\dgr(\br')
\Phi(\br-\br') \hat\psi(\br') \hat\psi(\br) \; d\br d\br' \; ,
\ee
where $U = U({\bf r},t)$ is an external field, $\Phi({\bf r})$ is an interaction 
potential and $\hat{\psi}({\bf r})$ are field operators obeying Bose commutation 
relations. In the presence of a Bose condensate, the field operator acquires the 
Bogolubov shift \cite{Bogolubov_4,Bogolubov_5,Bogolubov_14}
\be
\label{9}
 \hat\psi(\br) = \eta(\br) +\psi_1(\br) \; ,
\ee
in which $\eta$ is the condensate function and $\psi_1$ is the field operator of 
uncondensed particles. Note that representation (\ref{9}) is not an approximation 
but an exact canonical transformation. The variables $\eta$ and $\psi_1$ are mutually 
orthogonal,
\be
\label{10}
 \int \eta^*(\br) \psi_1(\br) \; d\br = 0 \; .
\ee
The condensate function plays the role of an order parameter satisfying the equations
\be
\label{11}
\eta(\br) = \lgl \; \hat\psi(\br) \; \rgl \; , \qquad  
\lgl \; \psi_1(\br) \; \rgl = 0 \; .
\ee
 
The condensate function is normalized to the number of condensed particles,
\be
\label{12}
 N_0 = \int |\; \eta(\br) \; |^2 \; d\br \; .
\ee
And the number of uncondensed particles is given by the statistical average
\be
\label{13}
N_1 = \lgl \; \hat N_1 \; \rgl \; ; \qquad \hat N_1 \equiv 
\int \psi_1^\dgr(\br) \psi_1(\br) \; d\br
\ee
for the number-of-particle operator of uncondensed particles. Condition 
$\langle \psi_1 \rangle = 0$ in a more general form writes as the average
\be
\label{14}
 \lgl \; \hat\Lbd \; \rgl = 0
\ee
of the operator
$$
\hat\Lbd  \equiv \int \left[ \lbd(\br)\psi_1^\dgr(\br) + 
\lbd^*(\br)\psi_1(\br) \right] \; d\br \;  .
$$
The grand Hamiltonian, taking into account the normalization conditions (\ref{12}) 
and (\ref{13}) and average (\ref{14}), becomes
\be
\label{15}
 H = \hat H - \mu_0 N_0 - \mu_1 \hat N_1 - \hat\Lbd \;  ,
\ee
in which $\mu_0$, $\mu_1$, and $\lbd({\bf r})$ are the Lagrange multipliers 
guaranteeing the validity of conditions (\ref{12}), (\ref{13}), and (\ref{14}). 

The equations of motion can be written as the equation for the condensate function
\be
\label{16}
i \; \frac{\prt}{\prt t} \; \eta(\br,t) = \left\lgl \;
\frac{\dlt H}{\dlt\eta^*(\br,t)} \; \right\rgl
\ee
and for the operator of uncondensed particles
\be
\label{17}
 i \; \frac{\prt}{\prt t} \; \psi_1(\br,t) =  
\frac{\dlt H}{\dlt\psi_1^\dgr(\br,t)} \; .
\ee
The operator equation (\ref{17}) is equivalent to the Heisenberg equation of motion, 
since it can be proved \cite{Yukalov_3} that
\be
\label{18}
 \frac{\dlt H}{\dlt\psi_1^\dgr(\br,t)} = [\; \psi_1(\br,t) , \; H \; ] \; .
\ee
The system chemical potential takes the form
\be
\label{19}
\mu = \mu_0 n_0 + \mu_1 n_1 \qquad \left( n_0 \equiv \frac{N_0}{N} \; , ~~
 n_1 \equiv \frac{N_1}{N} \right) \; .
\ee

This general approach includes the particular case, when the chemical potentials 
$\mu_0$ and $\mu_1$ are equal, which can happen in some simple approximations shown 
below.

\section{Quasiclassical approximation}

At asymptotically small temperature and weak interactions, when
\be
\label{20}
 T \ra 0 \; , \qquad \rho\Phi_0 \ra 0 \; ; \qquad 
\Phi_0 \equiv \int \Phi(\br) \; d\br \;  ,
\ee
almost all the system is Bose-condensed, hence it is in a coherent state, which is 
the eigenstate of the field operator,
\be
\label{21}
 \hat\psi(\br,t) \; | \; \eta \; \rgl = \eta(\br,t) \; | \; \eta \; \rgl \; .
\ee
The condensate function is defined by the coherent field $\eta({\bf r},t)$, for which 
Eq. (\ref{16}) yields the nonlinear Schr\"{o}dinger equation 
\be
\label{22}
i \; \frac{\prt}{\prt t} \; \eta(\br,t) = H[\; \eta \; ] \; \eta(\br,t) \; ,
\ee
with the nonlinear Hamiltonian
\be
\label{23}
H[\; \eta \; ] = -\; \frac{\nabla^2}{2m} + U(\br,t) - \mu + 
\int \Phi(\br-\br') \; |\; \eta(\br',t)\; |^2 \; d\br' \;  .
\ee
Equation (\ref{22}) was, first, advanced by Bogolubov in 1949 \cite{Bogolubov_15} 
(see also \cite{Bogolubov_4,Bogolubov_5,Bogolubov_14}). 

For a uniform system, considering small deviations of the condensate function
\be
\label{24}
 \eta(\br,t) = \sqrt{\rho_0}\; + \; ue^{-i\ep t} \; + \; v^* e^{i\ep t} \;  ,
\ee
one obtains the Bogolubov spectrum
\be
\label{25}
 \ep_k = \sqrt{c_B^2 k^2 + \left( \frac{k^2}{2m} \right)^2} \qquad
\left(\; c_B \equiv \sqrt{\frac{\rho}{m} }\; \Phi_0 \; \right) \; ,
\ee
with the Bogolubov sound velocity $c_B$. 

In the quasiclassical approximation, the chemical potentials $\mu_0$ and $\mu_1$ 
coincide, being equal to $\mu = \rho \Phi_0$.

\section{Bogolubov approximation}

This approximation is applicable for temperatures much lower than the 
condensation temperature, $T \ll T_c$, and weak interactions, such that the 
characteristic interaction energy be much lower than the characteristic kinetic 
energy $E_{kin} \sim 1/2m a^2$, where $a$ is mean interparticle distance. The 
latter condition gives the inequality
\be
\label{26}
  m\rho^{1/3} | \; \Phi_0 \; | \ll 1 \; .
\ee

In the Bogolubov approximation \cite{Bogolubov_4,Bogolubov_5}, one neglects in 
the Hamiltonian the fourth-order operator terms like $\psi_1^\dgr\psi_1^\dgr\psi_1\psi_1$. 
As a result, one gets the spectrum of collective excitations (\ref{25}) and the same 
chemical potentials $\mu_0=\mu_1=\mu=\rho\Phi_0$. But not all particles are condensed, 
even at zero temperature, since interactions deplete the condensate, whose density 
becomes
\be
\label{27}
 \rho_B  = \rho \left( 1 \; - \; \frac{8}{3\sqrt{\pi} }\; \gm^{3/2} \right) \; ,
\ee
where 
$$
\gm \equiv \rho^{1/3} a_s = \frac{m}{4\pi} \; \rho^{1/3}\Phi_0
$$
is the dimensionless gas parameter, with $a_s$ being scattering length.

\section{Hartree-Fock-Bogolubov approximation}

In the Hartree-Fock-Bogolubov approximation, the fourth-order operator terms are 
decoupled into second-order terms, so that for the fourth-order correlators one gets
\be
\label{28}
 \lgl \; \psi^\dgr_1 \psi^\dgr_1 \psi_1 \psi_1 \; \rgl = 
\lgl \; \psi^\dgr_1 \psi_1 \; \rgl  \lgl \; \psi^\dgr_1 \psi_1 \; \rgl + 
\lgl \; \psi^\dgr_1 \psi_1 \; \rgl  \lgl \; \psi^\dgr_1 \psi_1 \; \rgl +
\lgl \; \psi^\dgr_1 \psi^\dgr_1 \; \rgl  \lgl \; \psi_1 \psi_1 \; \rgl \; .
\ee
For a uniform system, the self-energies at zero variables become 
$$
\Sigma_{11}(0,) = (\rho + \rho_0)\Phi_0 + 
\int \Phi(\br) \lgl \; \psi^\dgr_1(0) \psi_1(\br) \; \rgl \; d\br \; ,
$$
\be
\label{29}
\Sigma_{12}(0,) = \rho_0\Phi_0 + 
\int \Phi(\br) \lgl \; \psi_1(0) \psi_1(\br) \; \rgl \; d\br \; .
\ee 
This yields the chemical potentials
$$
\mu_0 = \rho\Phi_0 + \int \Phi(\br) ( \; \lgl \; \psi^\dgr_1(0) \psi_1(\br)\; \rgl
+ \lgl \; \psi_1(0) \psi_1(\br)\; \rgl \; ) d\br \; , 
$$
\be
\label{30}
\mu_1 = \rho\Phi_0 + \int \Phi(\br) ( \; \lgl \; \psi^\dgr_1(0) \psi_1(\br)\; \rgl
- \lgl \; \psi_1(0) \psi_1(\br)\; \rgl \; ) d\br \; ,
\ee
that are evidently different \cite{Yukalov_3,Yukalov_10}. 

For a dilute gas, it is possible to use the local interaction potential
\be
\label{31}
\Phi(\br) = \Phi_0 \dlt(\br) \qquad 
\left( \Phi_0 = 4\pi\; \frac{a_s}{m} \right) \;  .
\ee
This gives the spectrum of collective excitations
\be
\label{32}
\ep_k = \sqrt{c^2k^2 + \left( \frac{k^2}{2m}\right)^2}   
\ee
that looks similar to the Bogolubov spectrum (\ref{25}), however with the sound 
velocity $c$ defined by the equation
\be
\label{33}
mc^2 = \rho\Phi_0 ( n_0 + \sgm )  \; ,
\ee
where $n_0 \equiv \rho_0/\rho$ and $\sigma$ is the anomalous average
\be
\label{34}
 \sgm = \frac{1}{\rho} \; \lgl \; \psi_1(\br) \psi_1(\br) \; \rgl \; .
\ee
In dimensionless variables
\be
\label{35}
s \equiv \frac{mc}{\rho^{1/3} } \; , \qquad \gm \equiv \rho^{1/3} a_s \; ,
\ee
the equation for the sound velocity reads as
\be
\label{36}
 s^2 = 4\pi\gm ( n_0 + \sgm ) \; .
\ee
 
The anomalous average can be represented as the sum of two terms, $\sgm=\sgm_0+\sgm_T$, 
where the first term is the value of $\sgm$ at zero temperature,
\be
\label{37}
 \sgm_0 = -\; 
\frac{mc^2}{2\rho} \int \frac{1}{\ep_k} \; \frac{d\bk}{(2\pi)^3} \;  .
\ee
This term displays an ultraviolet divergence and requires a regularization. We employ 
the dimensional regularization that is asymptotically exact for small interactions, 
when $\Phi_0\ra 0$. After accomplishing the dimensional regularization, it is necessary 
to analytically continue the obtained expression to finite interactions. The analytical 
continuation can be done by means of an iterative procedure \cite{Yukalov_16,Yukalov_17} 
starting from the ideal-gas limit, where $\sgm_0\ra 0$ for $\Phi_0\ra 0$ and respecting 
the symmetry-restoration limit, when $\sgm_0\ra 0$ for $n_0 \ra 0$. This procedure
results in the expression
\be
\label{38}
 \sgm_0 =  \frac{s_B^3}{\pi^2} \left(\; n_0 + 
\frac{s_B^3}{\pi^2} \; \sqrt{n_0} \;\right)^{1/2} \; ,
\ee
with the dimensionless Bogolubov sound velocity
$$
 s_B \equiv \frac{mc_B}{\rho^{1/3} } = \sqrt{4\pi\gm } \; .
$$
Using the definition of the gas parameter $\gamma$, we find
\be
\label{39}
  \sgm_0 =  \frac{8}{\sqrt{\pi} }\; \gm^{3/2} \; \left(\; n_0 + 
\frac{8}{\sqrt{\pi} } \; \gm^{3/2} \; \sqrt{n_0} \;\right)^{1/2} \;  .
\ee

\section{Numerical calculations}

The theory presented in the previous section has been applied for calculating the 
condensate density $n_0$ and superfluid faction $n_s$ in uniform and harmonically 
trapped systems \cite{Yukalov_16,Yukalov_18,Yukalov_19}. The results are shown in 
Figs. 1 to 5. In Fig. 1, the condensate fraction $n_0$, for a uniform system, as a 
function of temperature, in the vicinity of the condensation temperature, demonstrates 
that the Bose-Einstein condensation transition is of second order for any interaction 
strength, as it should be. Temperature here is measured in units of $m/ \rho^{2/3}$, 
and the interaction strength is characterized by the gas parameter $\gm$. The 
superfluid fraction $n_s$ appears at the same critical point, as is seen in Fig. 2. 
In  Fig. 3, we study the calculated condensate density $n_0$, for a uniform system 
at zero temperature as a function of the gas parameter and compare it with the Monte 
Carlo simulations \cite{Rossi_20} and the Bogolubov approximation $n_B$. In Fig 4, we 
show the effect of local condensate depletion at trap center occurring for the gas 
parameter $\gm>0.3$. And Fig. 5 shows the average condensate density in a harmonic trap 
as a function of the gas parameter, compared with the Bogolubov approximation 
\cite{Javanainen_21} and the Monte Carlo simulations \cite{DuBois_22}. The calculated 
results, based on the Hartree-Fock-Bogolubov approximation, are in good agreement with 
experiments and Monte Carlo simulations, contrary to the Bogolubov approximation valid 
only for weak interactions, where $\gm<0.1$.

\vskip 1cm

\begin{figure}[ht]
\centerline{
\includegraphics[width=6.7cm]{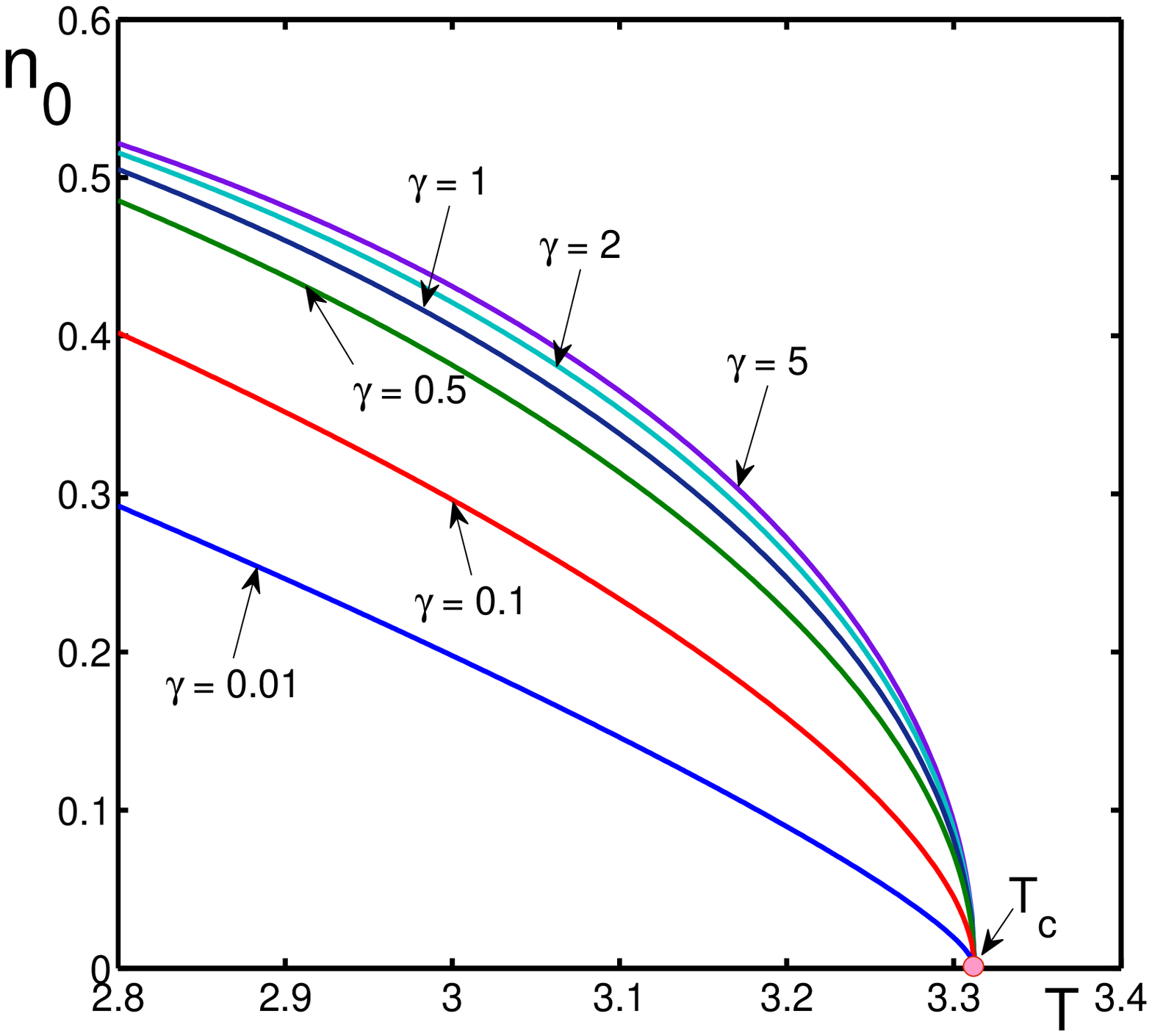} }
\caption{Condensate fraction $n_0$ for a uniform system as a function of dimensionless 
temperature (in units of $\rho^{2/3}/m$), for different gas parameters, in the vicinity 
of the Bose-Einstein condensation.}
\label{fig:Fig.1}
\end{figure}

\vskip 1cm

\begin{figure}[ht]
\centerline{
\includegraphics[width=6.7cm]{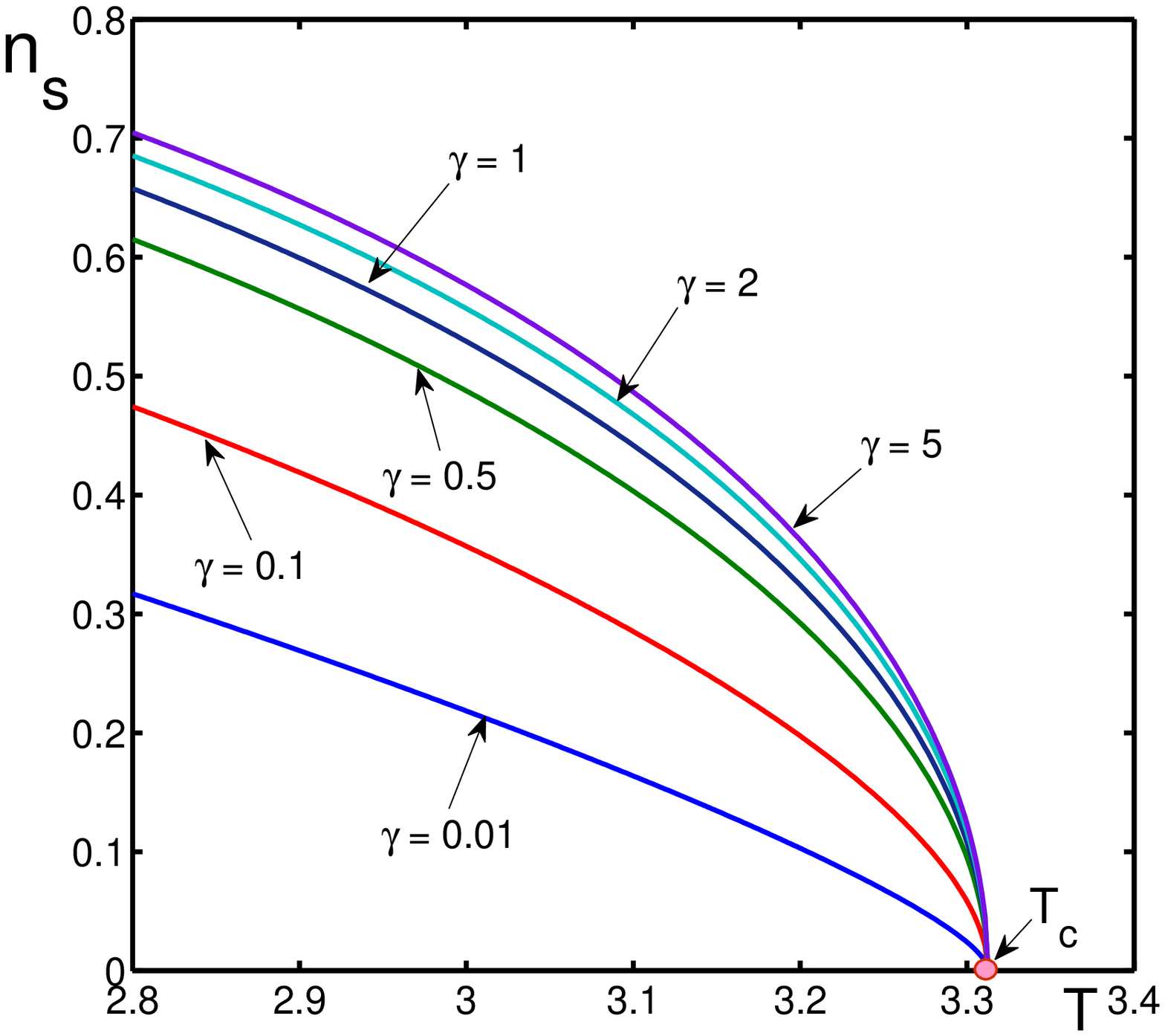} }
\caption{Superfluid fraction $n_s$ for a uniform system as a function of dimensionless 
temperature (in units of $\rho^{2/3}/m$), for different gas parameters, in the vicinity 
of the Bose-Einstein condensation.}
\label{fig:Fig.2}
\end{figure}

\vskip 1cm

\begin{figure}[ht]
\centerline{
\includegraphics[width=6.7cm]{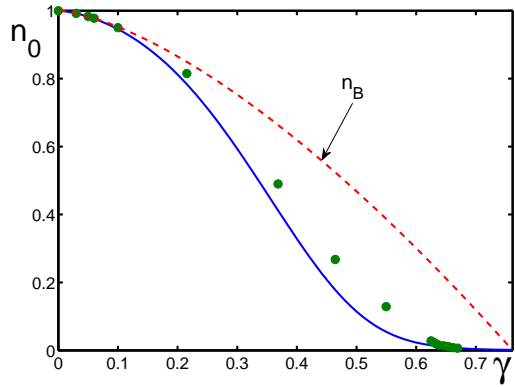} }
\caption{Condensate fraction at zero temperature for a uniform system as a function 
of the gas parameter (solid line) compared with the Monte Carlo simulations (dots) 
and the Bogolubov approximation $n_B$.}
\label{fig:Fig.3}
\end{figure}

\vskip 1cm

\begin{figure}[ht]
\centerline{\hbox{
\includegraphics[width=6.7cm]{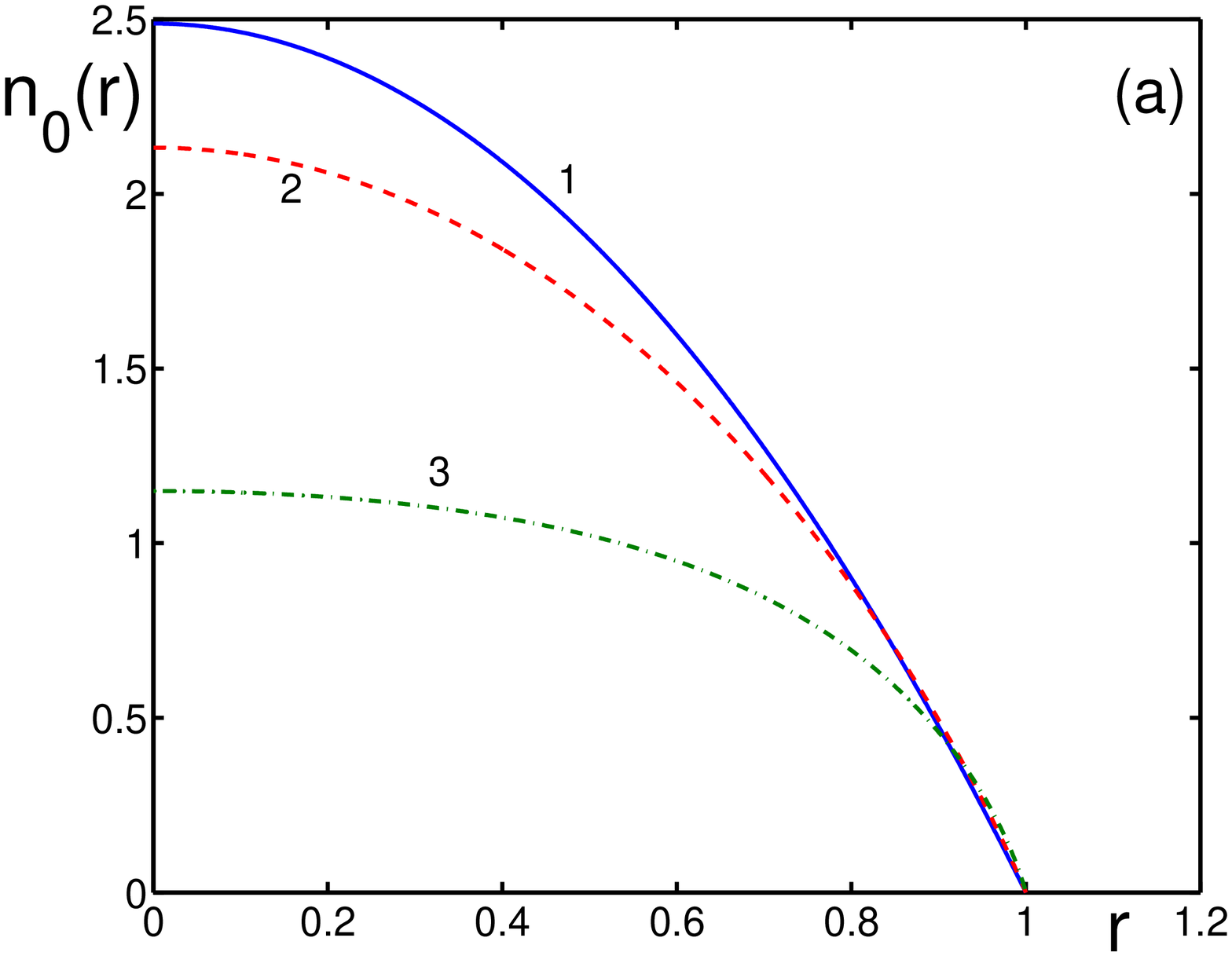} \hspace{1cm}
\includegraphics[width=6.7cm]{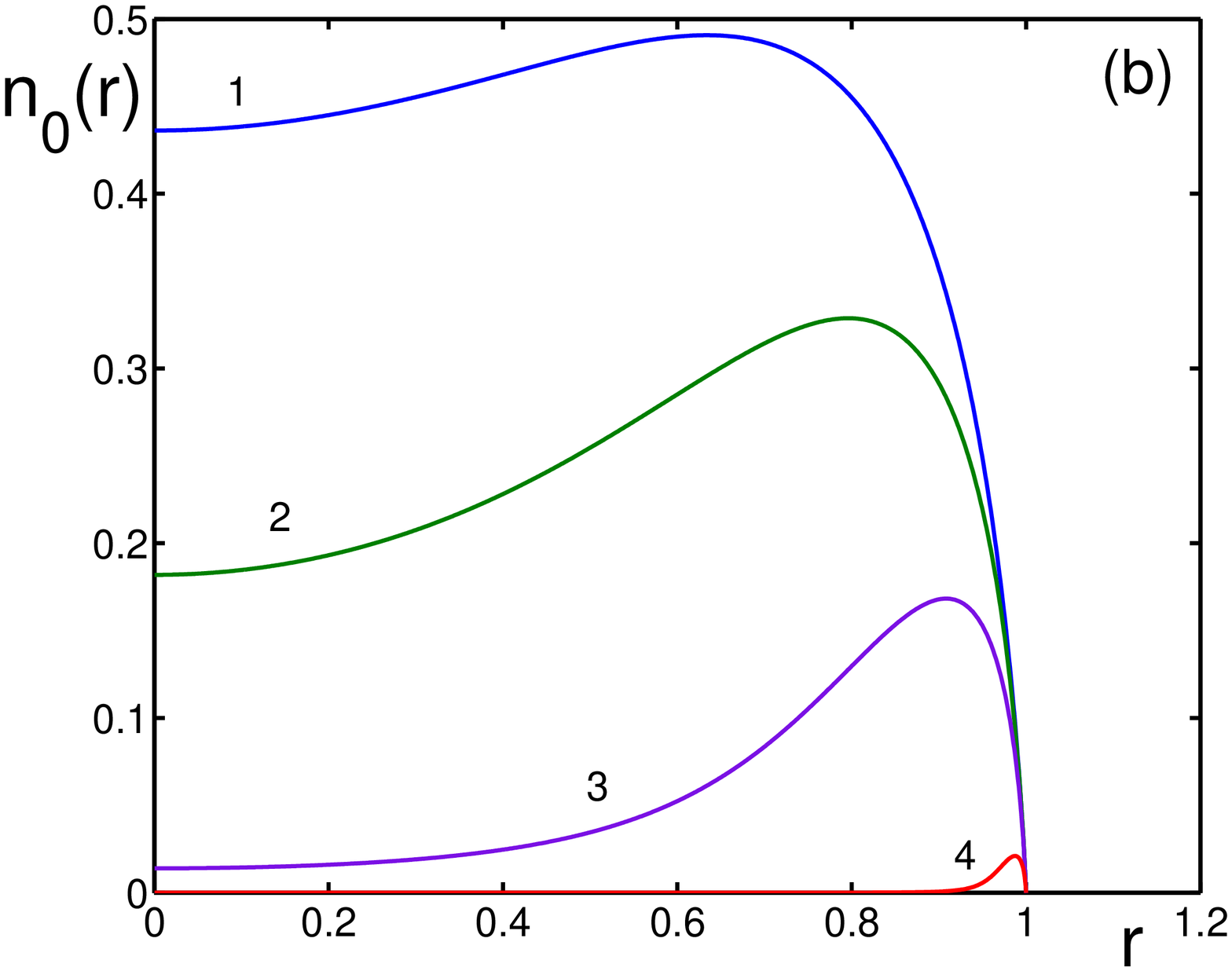} }   }
\caption{Spatial dependence of the condensate fraction in a trap at zero temperature, 
for different gas parameters. (a) No local trap-center depletion for small $\gm<0.3$: 
with $\gm=0.01$ (line 1), $\gm=0.1$ (line 2), and $\gm=0.25$ (line 3). (b) Effect of 
trap-center depletion for $\gm>0.3$: with $\gm=0.35$ (line 1), $\gm=0.4$ (line 2), 
$\gm=0.5$ (line 3), and $\gm=1$ (line 4).   
}
\label{fig:Fig.4}
\end{figure}

\vskip 1cm

\begin{figure}[ht]
\centerline{
\includegraphics[width=6.7cm]{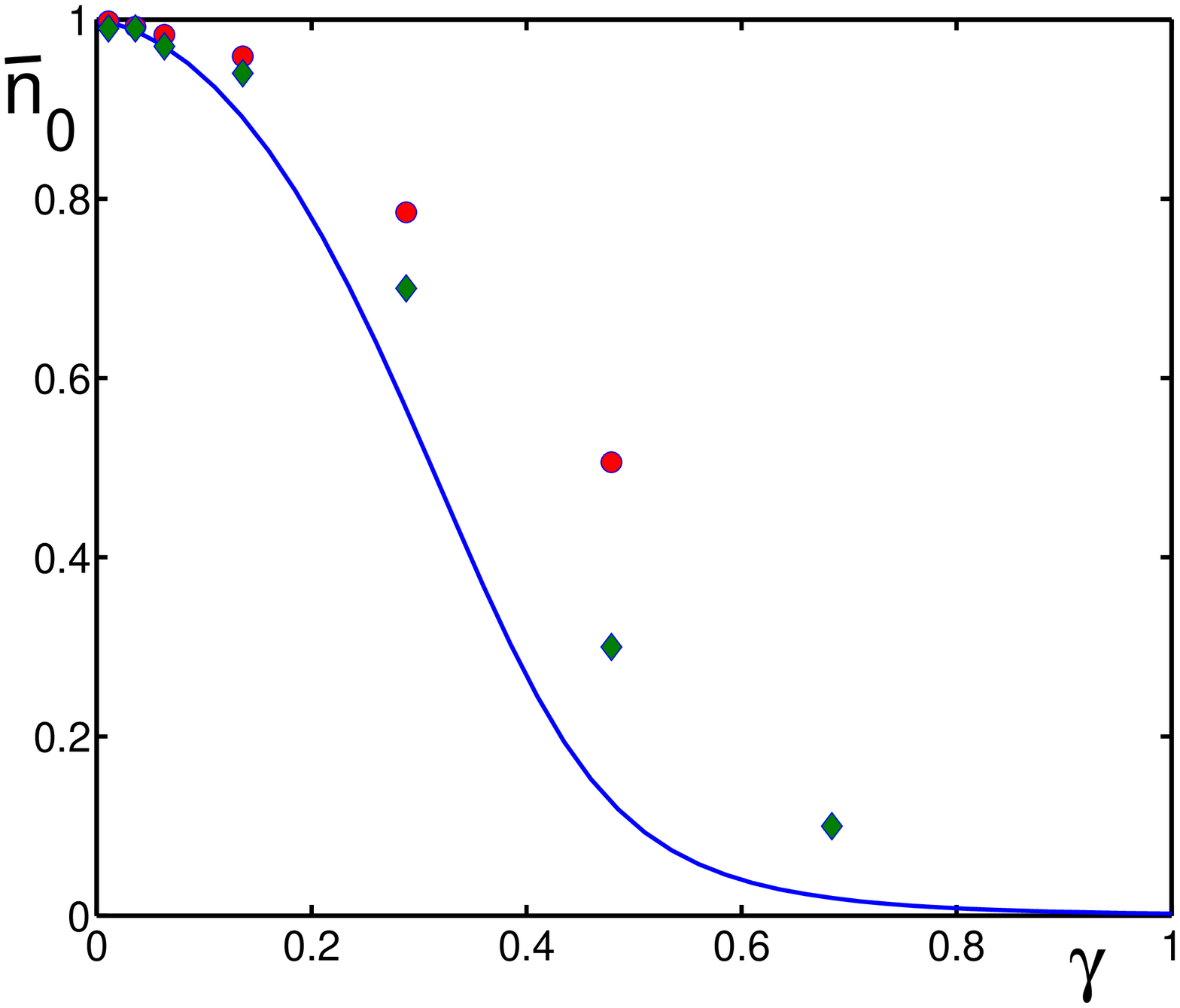} }
\caption{Mean condensate fraction in a trap at zero temperature, as a function of 
the gas parameter (solid line), compared with the Bogolubov approximation (fat dots) 
and Monte Carlo simulations (diamonds). 
}
\label{fig:Fig.5}
\end{figure}


\vskip 1cm

\begin{thebibliography}{99}

\bibitem{Lieb_1}
E. H. Lieb, R. Seiringer, J. P. Solovej, and J. Yngvason,
{\it The Mathematics of the Bose Gas and Its Condensation}
(Birkhauser, Basel, 2005).

\bibitem{Yukalov_2}
V. I. Yukalov,
Laser Phys. Lett. {\bf 4}, 632–647 (2007).

\bibitem{Yukalov_3}
V. I. Yukalov, 
Phys. Part. Nucl. {\bf 42}, 460--513 (2011). 

\bibitem{Bogolubov_4}
N. N. Bogolubov, 
{\it Lectures on Quantum Statistics} (Gordon and Breach, New York, 1967), Vol. 1.

\bibitem{Bogolubov_5}
N. N. Bogolubov, 
{\it Lectures on Quantum Statistics} (Gordon and Breach, New York, 1970), Vol. 2.

\bibitem{Ginibre_6}
J. Ginibre, 
Commun. Math. Phys. {\bf 8}, 26--51 (1968).

\bibitem{Penrose_7}
O. Penrose and L. Onsager, 
Phys. Rev. {\bf 104}, 576--584 (1956).

\bibitem{Hugenholtz_8}
N. M. Hugenholtz and D. Pines, 
Phys. Rev. {\bf 116}, 489--506 (1959).

\bibitem{Nepomnyashchii_9}
A. A. Nepomnyashchii and Y. A. Nepomnyashchii,
JETP Lett. {\bf 21}, 3--6 (1975).

\bibitem{Hohenberg_9}
P. C. Hohenberg and P. C. Martin, 
Ann. Phys. (N.Y.) {\bf 34}, 291--359 (1965).

\bibitem{Yukalov_10}
V. I. Yukalov,
Ann. Phys. (N.Y.) {\bf 323}, 461--499 (2008)

\bibitem{Yukalov_11}
V. I. Yukalov,
Laser Phys. {\bf 26}, 062001 (2016). 

\bibitem{Yukalov_12}
V. I. Yukalov,
Phys. Rev. A {\bf 72}, 033608 (2005).

\bibitem{Yukalov_13}
V. I. Yukalov,
Symmetry {\bf 11}, 603 (2019).

\bibitem{Bogolubov_14}
N. N. Bogolubov, 
{\it Quantum Statistical Mechanics} (World Scientific, Singapore, 2015).

\bibitem{Bogolubov_15}
N. N. Bogolubov, 
{\it Lectures on Quantum Statistics} (Ryadyanska Shkola, Kiev, 1949).

\bibitem{Yukalov_16}
V. I. Yukalov and E. P. Yukalova,
Phys. Rev. A {\bf 90}, 013627 (2014).

\bibitem{Yukalov_17}
V. I. Yukalov and E. P. Yukalova,
Laser Phys. Lett. {\bf 16}, 065501 (2019).

\bibitem{Yukalov_18}
V. I. Yukalov and E. P. Yukalova,
J. Phys. B {\bf 47}, 095302 (2014).

\bibitem{Yukalov_19}
V. I. Yukalov and E. P. Yukalova,
J. Phys. B {\bf 51}, 085301 (2018).

\bibitem{Rossi_20}
M. Rossi and L. Salasnich,
Phys. Rev. A {\bf 88}, 053617 (2013).

\bibitem{Javanainen_21}
J. Javanainen,
Phys. Rev. A {\bf 54}, 3722 (1996).

\bibitem{DuBois_22}
J. L. DuBois and H. R. Glyde,
Phys. Rev. A {\bf 68}, 033602 (2003).
\end{thebibliography}
\end{document}